\documentclass[onecolumn,amsmath,amssymb]{qm2008_abs}
\usepackage{graphicx}
\usepackage{dcolumn}
\usepackage{bm}
\usepackage{epsfig}
\begin{document}

\title{{Photons from Jet - Plasma Interaction in collisional energy loss
scenario}}
\bigskip
\bigskip
\author{\large Lusaka Bhattacharya$^1$}
\email{lusaka.bhattacharya@saha.ac.in}
\author{\large Pradip k Roy$^2$}
\email{pradipk.roy@saha.ac.in}
\affiliation{$^1$Saha Institute of Nuclear Physics, Kolkata - 700064, INDIA}

\bigskip
\bigskip

\begin{abstract}
\leftskip1.0cm
\rightskip1.0cm
We calculate photons from jet plasma interaction in a collisional energy loss
scenario. It is shown that the Phenix photon data is well reproduced when
photons from initial hard collisions are taken into account.
\end{abstract}

\maketitle

\section{Introduction}

Study of direct photon and dilepton spectra emanating from hot and dense
hadronic matter formed in ultra-relativistic heavy ion collisions is a
field of considerable
current interest. Electromagnetic probes have been proposed to be one
of the most promising tools to characterize the initial state of
the collisions~\cite{jpr}. Because of the very nature of their
interactions
with the constituents of the system they tend to leave the system without
much change of their energy and momentum. In fact, photons (dilepton as well)
can be used to determine the initial temperature, or equivalently the
equilibration time. These are related to the final multiplicity of
the produced hadrons in relativistic heavy ion collisions. By comparing
the initial temperature with the transition
temperature from lattice QCD, one can infer whether Quark Gluon Plasma (QGP)
 is formed or not.

There are various sources of photons from relativistic heavy ion collisions:
(i) Thermal photons from QGP and hot hadronic matter (HHM),
(ii) hard photons ($A + B \,\rightarrow\,\gamma\,X$).
(iii) photons from decay of  $\pi^0 (\eta)\,\rightarrow\,\gamma\,\gamma$.
Hard photon yield can be reliably calculated using
perturbative quantum chromodynamics.

The last class of photon emission processes is the jet conversion
mechanism~\cite{dks} which occurs when a high energy jet interacts
with the medium constituents via annihilation and compton processes. It
might be noted that this phenomenon (for Compton process) has been
illustrated quite some time ago~\cite{pkrnpa} in the context of
estimating photons from equilibrating scenario where, because of the
larger cross-section, gluons equilibrate among themselves providing a heat
bath to the incoming quark-jet. A comparison of the non-equilibrium
photons (equivalent to jet-conversion) with the direct photons (thermal)
shows that this contribution remains dominant for photons with $p_T$
upto 6 GeV. However, while evaluating jet-photon the assumption made in
Ref.~\cite{dks} that the largest contribution to photons corresponds
to $p_{\gamma} \sim p_q (p_{\bar q})$ which amounts to saying that the
annihilating quark (anti-quark) directly converts to a photon. Moreover,
before annihilating the quark jet loses energy in the scattering with the
particles in the thermal bath.

The partonic energy loss due to collisional processes was
revisited in~\cite{abhee05} and its importance was demonstrated in the
context of RHIC in\cite{roy06}. The measurements of non-photonic
single electron data~\cite{phenixdil} show larger suppression than expected.
These electrons mainly come from heavy quark decay where the radiative
energy loss is suppressed due to dead cone effect. This observation
has led to re-thinking the importance of collisional energy loss
both for heavy as well as light quarks.

In this work we calculate the photon yield from jet-plasma interaction
using exact expression for photon rate.
We also include the jet energy loss
in the jet-plasma interaction. In view of the controversy over the
relative importance between $2\to2$ and $2\to3$ processes we restrict
ourselves to the collisional energy loss only. In the photon production
rate (from jet-plasma interaction) one of the collision partners
 is in equilibrium and the other (the jet) is
assumed to execute Brownian motion in the heat bath consisting of
quarks ( anti-quarks) and gluons. Furthermore, the collisional energy loss
is dominated by small angle scattering. Under such scenario the evolution
of the jet phase space distribution is governed by Fokker-Planck (FP) equation
where the collision integral is approximated by appropriately defined
drag and diffusion coefficients.

\section{Theory}
The lowest order processes for photon emission from QGP are the
Compton ($q ({\bar q})\,g\,\rightarrow\,q ({\bar q})\,
\gamma$) and annihilation ($q\,{\bar q}\,\rightarrow\,g\,\gamma$)
processes.
The total cross-section diverges
since the differential cross-section suffers singularity at $t$ and/or $u=0$.
These singularities have to be shielded by thermal effects in order to
obtain infrared safe calculations. It has been argued in Ref.~\cite{kajruus}
that the intermediate quark acquires a thermal mass in the medium, whereas
the hard thermal loop (HTL) approach of Ref.~\cite{Brapi} shows that very
soft modes are suppressed in a medium leading to a natural cut-off
$k_c \sim gT$.

We, from the very beginning, assume the singularities can be shielded by the
introduction of thermal masses for the participating partons.
The differential cross-sections for Compton and
annihilation processes are taken from Ref.~\cite{wong}.
The static photon rate in
$1+2\rightarrow 3+\gamma$ can be
written as~\cite{jpr}
\begin{eqnarray}
\frac {dN}{d^4xd^2p_T dy}&=&\frac{\mathcal {N}_i}{16(2\pi)^7 E_\gamma}\int ds dt
|\mathcal{M}|^2 \int dE_1 dE_2 
\frac{f_1(E_1) f_2(E_2)(1+f_3(E_3))}{\sqrt{a{E_2}^2+2bE_2+c}}\\
\rm where \nonumber\\
a&=&-(s+t-{m_2}^2-{m_3}^2)^2\nonumber\\ 
b&=&E_1(s+t-{m_2}^2-{m_3}^2)({m_2}^2-t)+E[(s+t-{m_2}^2-{m_3}^2)\nonumber\\
&\times&(s-{m_1}^2-{m_2}^2)-2{m_1}^2({m_2}^2-t)]\nonumber\\
c&=&{E_1}^2({m_2}^2-t)^2-2E_1E[2{m_2}^2(s+t-{m_2}^2-{m_3}^2)\nonumber\\
&-&({m_2}^2-t)(s-{m_1}^2-{m_2}^2)]-E^2[(s-{m_1}^2-{m_2}^2)^2-4{m_1}^2{m_2}^2]
\nonumber\\
&-&(s+t-{m_2}^2-{m_3}^2)({m_2}^2-t)(s-{m_1}^2-{m_2}^2)\nonumber\\
&+&{m_2}^2(s+t-{m_2}^2-{m_3}^2)^2+{m_1}^2({m_1}^2-t)^2\nonumber\\
E_{1,min}&=&\frac{s+t-{m_2}^2-{m_3}^2}{4E}
+\frac{E{m_1}^2}{s+t-{m_2}^2-{m_3}^2} \nonumber\\
E_{2,min}&=&\frac{E{m_2}^2}{{m_2}^2-t}+\frac{{m_2}^2-t}{4E}\nonumber\\
E_{2,max}&=&-\frac{b}{a}+\frac{\sqrt{b^2-ac}}{a}\nonumber\\
E_3&=& E_1+E_2-E_{\gamma} \nonumber
\end{eqnarray}
$ f_1 (E_1)$ , $f_{2}(E_2)$ and $ f_3(E_3)$ are the distribution
functions of parton respectively. $\mathcal {M}$ represents the amplitude for
compton or annihilation process. $\mathcal {N}_i$ is the overall degeneracy
factor. For compton process $\mathcal {N}_i=320/3$ and for annihilation
process $\mathcal {N}_i=20$ when summing over u and d quarks.

As mentioned in the earlier the jet phase space distribution
can be obtained by solving FP equation which reads as,
\begin{eqnarray}
\left (\frac{\partial}{\partial t}
-\frac{p_\parallel}{t}\frac{\partial}{\partial p_\parallel}\right )f({\bf p},t)
&=&\frac{\partial}{\partial p_i}[p_i\eta f({\bf p},t)]
+\frac{1}{2}\frac{\partial^2}{\partial p_\parallel^2 }
[{B_\parallel}({\bf p})f({\bf p},t)]
+\frac{1}{2}\frac{\partial^2}{\partial p_\perp^2}[{B_\perp}f({\bf p},t)]
\label{fpexp}
\end{eqnarray}
where the second term on the left hand side arises due to
expansion~\cite{baym}. In Eqn.~(\ref{fpexp})
$f({\bf p},t)$ represents the non-equilibrium distribution of the
partons under study, $\eta=(1/E)dE/dx$, denotes drag coefficient,
$B_\parallel=d\langle(\Delta p_\parallel)^2\rangle/dt$,
$B_\perp=d\langle(\Delta p_\perp)^2\rangle/dt$,
represent diffusion constants along parallel and perpendicular directions
of the propagating partons.

The transport coefficients, $\eta$, $B_\parallel$ and $B_\perp$
appeared in Eqn.(~\ref{fpexp}) can be calculated from the
following expressions:
\begin{eqnarray}
\frac{dE}{dx}=\frac{\nu} {(2\pi)^5}
\int\frac{d^3kd^3qd\omega}{2k2k^\prime 2p 2p^\prime}
~\delta(\omega-{\bf v_p\cdot q})\delta(\omega-{\bf v_k\cdot q})
\langle {\cal M} \rangle_{t\rightarrow 0}^2
f(|{\bf k}|)\left[1+f(|{\bf k}+{\bf q}|)\right]\omega
\end{eqnarray}
\begin{eqnarray}
B_{\perp,\parallel}=\frac{\nu }{(2\pi)^5}
\int\frac{d^3kd^3qd\omega}{2k2k^\prime 2p 2p^\prime}
~\delta(\omega-{\bf v_p\cdot q})\delta(\omega-{\bf v_k\cdot q})
\langle {\cal M} \rangle_{t\rightarrow 0}^2
f(|{\bf k}|)\left[1+f(|{\bf k}+{\bf q}|)\right]
q_{\perp,\parallel}^2 \nonumber\\
\equiv 2\,D_{\perp,\parallel}.
\end{eqnarray}
in the small angle limit~\cite{abhee05,roy06}.
Here $f(|{\bf k}|,t)$ denotes the thermal distributions for the
quarks (Fermi-Dirac) or gluons (Bose-Einstein).
The  matrix elements required to calculate the transports coefficients include
diagrams involving exchange of massless gluons which render
$dE/dx$ and $B_{\parallel,\perp}$ infrared divergent. Such
divergences can naturally be cured by using the hard thermal
loop (HTL)~\cite{Brapi} corrected propagator for the gluons, i.e.
the divergence is shielded by plasma effects.
For jet with energy $E >> T$ (see ~\cite{abhee05} for details) the energy
loss is given by
\begin{equation}
\frac{dE}{dx} \sim \alpha_s^2\,T^2\,C_R \ln\frac{E}{g^2T}
\end{equation}
Having known the drag and diffusion, we solve the FP equation using
Green's function techniques with the initial condition
\begin{equation}
P(\vec{p},t=t_i|\vec{p_0},t_i) = \delta^{(3)}(\vec{p}-\vec{p_0})
\end{equation}
in Bjorken expansion scenario~\cite{bj} along the line of
Refs.~\cite{moore05,rapp}.

The solution with an arbitrary initial momentum distribution
can now be written as~\cite{moore05,rapp},
\begin{equation}
E\frac{dN}{d^3p^f}|_{y=0} = \int\,d^3p_0^f\,
P(p^f,t|p_0^f,t_i)
E_0\frac{dN}{d^3p_0^f}|_{y_0=0}\,
\end{equation}
where $f$ stands for any parton species. We use the initial parton $p_T$
distributions (at the formation time $t_i$) taken from~\cite{muller}:
\begin{eqnarray}
\frac{dN}{d^2p_{0T}^fdy_0}|_{y_0=0}=
\frac{N_0}{(1+\frac{p_{0T}^f}{\beta})^\alpha},
\end{eqnarray}
In order to obtain the space-time integrated rate we first note that
the phase space distribution function for the incoming jet in the mid rapidity
region is given by(see Ref.~\cite{moore05} for details)
\begin{eqnarray}
f_{jet}(\vec r,\vec p,t^{\prime})|_{y=0}=
\frac{(2\pi)^3\mathcal{P}(\vec {w_r})~t_i}{g_q \sqrt{{t_i}^2-{z_0}^2}}
\frac{1}{p_{T}}\times\frac{dN}{d^2{p_{0T}}^fdy}(p,t^{\prime})\delta(z_0)
\end{eqnarray}
where $t_i$ is the jet formation tome. $z_0$ is the jet formation position in
the direction of QGP expansion and $\mathcal{P}(\vec {w_r})$ is the initial jet
production probability distribution at the initial radial position $\vec {w_r}$
in the plane $z_0=0$.
We assume the plasma expands only longitudinally. Thus
using $d^4x=rdrdt^{\prime}d\eta$ we obtain the transverse momentum
distribution as follows:
\begin{eqnarray}
\frac{dN^{\gamma}}{d^2p_Tdy}&=&\int d^4x ~ \frac{dN}{d^4xd^2p_Tdy} \nonumber\\
&=&\frac{(2\pi)^3}{g_q}{\int_{t_i}}^{t_c}dt^{\prime}
{\int_0}^R rdr \int d\phi\mathcal{P}(\vec {w_r})
\times\frac{N}{16(2\pi)^7E_{\gamma}}\int dsdt |\mathcal{M}|^2\nonumber\\
&\times&\int {dE_1 dE_2}\frac{ f_q(E_1,r,t^{\prime})f_{\bar q}(E_2)
(1+f_g(E_3))}{\sqrt{a{E_2}^2+2bE_2+c}}\,\,\nonumber
\end{eqnarray}
$\phi$ dependence occurs only in $\mathcal{P}(\vec {w_r})$. So the $\phi$
integration can be done analytically as in Ref.~\cite{moore05}.
The temperature profile is
taken from Ref.~\cite{moore05}.

\section{Results}

We plot the transverse momentum distribution of quarks in Fig.1 for different
times. It is seen that as the time increases the quark stays longer in
the medium losing more energy. As a result there is depletion in the 
distribution.
\begin{figure}
\hspace{-1.4cm}
\begin{minipage}[t]{7.20cm}
\includegraphics[scale=0.3,angle=0]{RHIC_ptdist.eps}
\caption{$p_T$ distribution of quarks at RHIC energies at initial 
temperature $T_i=446$MeV.}
\label{fig1}
\end{minipage}
\hspace{0.50cm}
\begin{minipage}[t]{6.0cm}
\includegraphics[scale=0.3,angle=0]{RHIC_31.eps}
\caption{$p_T$ distribution of photons at RHIC energies at initial 
temperature $T_i=446$MeV.}
\label{fig2}
\end{minipage}
\end{figure}
In Fig.2 we show $p_T$
distribution of photons from various processes which contribute at
this high $p_T$ range.The red (blue) curve denotes the  photon yield from 
jet-plasma interaction with (without) energy loss. The magenta 
represents the total yield compared with the Phenix measurements of 
photon data~\protect\cite{phenix} .It is observed that due to the 
inclusion of energy loss in the jet-plasma interaction the yield is depleted.
The total photon yield consists of jet-photon, photons from initial
hard collisions, bremsstrahlung photons, and thermal photons. It
is seen that Phenix photon data is well reproduced in our model.
At high $p_T$ region the data is marginally reproduced. The
reason behind this is the following.
For high $p_T$ photon the incoming jet must have high energy where
the radiative loss starts to dominate. Inclusion of this mechanism will
further deplete the photon yield at high $p_T$ reducing the total yield.

\section{Summary}

We have calculated the transverse momentum distribution of
photons from jet plasma interaction in a collisional energy loss
scenario. Our results for jet-plasma interaction is similar to what
is obtained in Ref.~\cite{dks} and ~\cite{fries}. Phenix photon data
have been contrasted with the present calculation and it is seen that
the data is reproduced quite well. We did not include the radiative energy
loss in our calculation as we think that in the measured domain at RHIC
collisional energy loss plays a vital role.

\end{document}